\documentclass[tikz,fleqn,usenatbib]{aa}

\usepackage{lastpage}
\usepackage{graphicx}
\usepackage{dcolumn}
\usepackage{bm}
\usepackage{natbib}
\usepackage{multirow}
\usepackage{url}
\usepackage{color}
\usepackage{xcolor}
\usepackage{tikz}
\usepackage{float}
\usepackage{cancel}
\usepackage{amssymb}
\usepackage{amsmath}
\usepackage{colortbl} 
\usepackage{breakurl}
\usepackage{comment}
\usepackage{subfig}
\usepackage{mathptmx}
\usepackage{txfonts}
\usepackage[T1]{fontenc}
\usepackage{ae,aecompl}
\usepackage{orcidlink}
\usepackage{academicons}
\usepackage{hyperref}
\usepackage{xcolor}
\usepackage{graphicx}
\usepackage{epstopdf}
\usepackage{amssymb}
\usepackage{latexsym}
\usepackage{tikz-cd}
\usepackage{natbib,twoopt}
\newcommand{\be}{\begin{equation}}

\newcommand{\mmax}{{\mathrm{max}}}
\newcommand{\vvir}{{\mathrm{vir}}}
\newcommand{\lp}{\left(}
\newcommand{\rp}{\right)}

\newcommand{\vr}{\textbf{r}}

\newcommand{\ee}{\end{equation}}

\newcommand{\vk}{\textbf{k}}

\usepackage{tikz}
\usetikzlibrary{matrix, fit, positioning,arrows.meta,arrows, shapes.geometric,backgrounds,shadows}
\tikzstyle{start}=[circle, rounded corners, minimum width=2cm, minimum height= 0cm, text centered, draw=black!3, fill=gray!25, text width = 2cm]
\tikzstyle{title}=[rectangle, rounded corners, minimum width=0.4cm, minimum height= 1cm, text centered, draw=black!5, fill=gray!0, text width = 3cm]
\tikzstyle{title_s}=[rectangle, rounded corners, minimum width=0.4cm, minimum height= 1cm, text centered, draw=black!0, fill=gray!0, text width=2.5cm]
\tikzstyle{ref_title}=[rectangle, rounded corners, minimum width=0.8cm, minimum height= 1cm, text centered, draw=black!50, fill=gray!2, text width = 2cm,drop shadow]
\tikzstyle{start_title}=[rectangle, rounded corners, minimum width=0.8cm, minimum height= 1cm, text centered, draw=black!50, fill=gray, text width = 2cm,drop shadow]
\tikzstyle{estart}=[circle, rounded corners, minimum width=0.1cm, minimum height= 0.1cm, text centered, draw=black!0, fill=red!0, text width = 1cm]

\tikzstyle{startstop}=[rectangle, rounded corners, minimum width=2.8cm, minimum height= 1cm, text centered, draw=black!6, fill=red!2, text width = 2.5cm,drop shadow]
\tikzstyle{box_s}=[rectangle, rounded corners, minimum width=3.5cm, minimum height= 1cm, text centered, draw=black!8, fill=red!5, text width = 4.5cm,drop shadow]
\tikzstyle{box_r}=[rectangle, rounded corners, minimum width=3.5cm, minimum height= 1cm, text centered, draw=black!8, fill=blue!5, text width = 4cm,drop shadow]
\tikzstyle{box_s_k}=[rectangle, rounded corners, minimum width=6cm, minimum height= 1cm, text centered, draw=black!6, fill=red!10, text width = 5cm,drop shadow]

\tikzstyle{startstop_nb}=[rectangle, rounded corners, minimum width=2.8cm, minimum height=0.8cm, text centered, draw=black, fill=gray!0, text width = 2cm]
\tikzstyle{io}=[trapezium, trapezium left angle = 70, trapezium right angle =110,minimum width=2cm, minimum height= 1cm, text centered, draw=black, fill=blue!0,drop shadow]
\tikzstyle{process}=[rectangle, rounded corners,minimum width=3cm, minimum height= 0.7cm, text centered, draw=black, fill=orange!0,drop shadow]

\tikzstyle{process_TITLE}=[rectangle, rounded corners, minimum width=1cm, minimum height= 0.4cm, text centered, draw=black, fill=gray!0,drop shadow]

\tikzstyle{nomen}=[rectangle, rounded corners, minimum width=4cm, minimum height= 1cm,  draw=black!10, fill=yellow!2]
\tikzstyle{process_short}=[rectangle, rounded corners,minimum width=1cm, minimum height=0.5cm, text centered, draw=black, fill=orange!0,drop shadow]
\tikzstyle{decision}=[diamond, rounded corners,minimum width=2cm, minimum height= 0.7cm, text centered, draw=black!15, fill=green!0,drop shadow]
\tikzstyle{question}=[diamond,rounded corners, minimum width=2.5cm, minimum height= 0.7cm, text centered, draw=black, fill=blue!0,drop shadow]
\tikzstyle{decisionfwd}=[diamond, rounded corners,minimum width=2cm, minimum height= 0.7cm, text centered, draw=black, fill=green!0, drop shadow]
\tikzstyle{decision_short}=[diamond, rounded corners,minimum width=1cm, minimum height= 1cm, text centered, draw=black!2, fill=green!5, drop shadow]

\tikzstyle{bam}=[diamond, rounded corners,minimum width=1cm, minimum height= 1cm, text centered, draw=black!15, fill=green!5, drop shadow]

\tikzstyle{decision_short_input}=[diamond, minimum width=0.8cm, minimum height=0.05cm, text centered, draw=black, fill=gray!0, drop shownshadow]
\tikzstyle{arrow}=[thick, ->, >= stealth,line width=0.05cm]
\tikzstyle{arrow_ic}=[thick, ->, >= stealth,line width=0.05cm, draw=black!40]

\tikzstyle{arrow_d}=[dotted, ->, >= stealth]
\tikzstyle{arrow_pi}=[dashed, ->, >= stealth, line width=0.03cm]
\tikzstyle{arrow_nl}=[dashed, -, >= stealth, line width=0.03cm]
\tikzstyle{arrow_nls}=[thick, -, >= stealth, line width=0.03cm]
\tikzstyle{arrow_nlsa}=[thick, ->, >= stealth, line width=0.03cm,color=blue]
\tikzstyle{arrow_new}=[dotted, ->,>= stealth, color=red, line width = 0.05cm]
\tikzstyle{arrow_prep}=[dashed, ->,>= stealth, color=blue!50, line width = 0.04cm]
\tikzstyle{arrow_cat}=[dotted, ->,>= stealth, color=green!80, line width = 0.05cm]
\tikzstyle{arrow_prep_nh}=[dashed, -,>= stealth, color=blue, line width = 0.04cm]
\begin{document}

\title{Secondary halo bias through cosmic time II: Reconstructing halo properties using clustering information}
\titlerunning{Balaguera-Antolínez and Montero-Dorta}
\author{
Andrés Balaguera-Antolínez $^{1,2}$\thanks{balaguera@iac.es}\orcidlink{0000-0001-5028-3035} and 
Antonio D. Montero-Dorta$^{3}$}
\institute{
 Instituto de Astrof\'{\i}sica de Canarias, s/n, E-38205, La Laguna, Tenerife, Spain \and
 Departamento de Astrof\'{\i}sica, Universidad de La Laguna, E-38206, La Laguna, Tenerife, Spain \and
 Departamento de Física, Universidad Técnica Federico Santa María, Casilla 110-V, Avda. España 1680, Valparaíso, Chile
  }
\authorrunning{Author}

\date{Received /Accepted}
\abstract{When constructing galaxy mock catalogs based on suites of dark matter halo catalogs generated with approximated, calibrated or machine-learning approaches, the assignment of intrinsic properties for such tracers is a step of paramount relevance, given that these can shape the abundance and spatial distribution of mock galaxies and galaxy clusters.}
{We explore the possibility to assign properties of dark matter halos within the context of calibrated/learning approaches, explicitly using clustering information.  The goal is to retrieve the correct signal of primary and secondary large-effective bias as a function of properties reconstructed solely based on phase-space properties of the halo distribution and dark matter density field.}
{The algorithm reconstructs a set halo properties (such as virial mass, maximum circular velocity, concentration and spin) constraint to reproduce both primary and secondary (or assembly) bias. The key ingredients of the algorithm are the implementation of individually-assigned large-scale effective bias, a multi-scale approach to account for halo exclusion and a hierarchical assignment of halo properties.}
{The method facilitates the assignment of halo properties aiming at replicating the large-scale effective bias, both primary and secondary. This improves over previous methods in the literature, especially at the high-mass end population.}
{We have designed an strategy to reconstruct the main properties of dark matter halos obtained by calibrated/learning algorithms, such that the one and two-point statistics (on large scales) replicates the signal from detailed $N$-body simulations. We encourage the application of this strategy (or the implementation of our algorithm) for the generation of mock catalogs of dark matter halos based on approximated methods.}
\maketitle

% *******************************************************************************************************************
% *******************************************************************************************************************
\section{Introduction}
Mock catalogs of galaxies and galaxy clusters are essential tools for the statistical analysis of galaxy redshift surveys (such as EUCLID \cite[][]{Euclid}, DESI \cite[][]{DESI}, J-PAS \citep{2014arXiv1403.5237B}, or the \emph{Nancy Grace Roman} space telescope \cite[][]{2015arXiv150303757S}).  Their usefulness lies not only in their capability to capture the main statistical properties of the spatial distribution of dark matter tracers in the Universe ($n-$ point statistics), but also the behavior of such properties as a function of a number of intrinsic properties (e.g., luminosity, stellar mass, color, star formation rate) which are key to understand the processes of galaxy cluster assembly and galaxy formation and evolution \citep[see e.g.,][]{2024arXiv240513495E}. 

Intrinsic galaxy and galaxy cluster properties in mock catalogs can be robustly derived from high resolution hydro-simulations \citep[e.g.,][]{2019ComAC...6....2N,2021A&A...651A.109D,2023MNRAS.522.3831F}, the cost of which typically scales with the size of the sample aimed at being reproduced, along with the details of the baryonic process involved, making this path demanding at replicating the vast cosmological volumes  currently surveyed. While this obstacle is rapidly being overcome \citep[see e.g.][]{2023MNRAS.526.4978S}, the true difficulty arises when thousand of realizations of such type of simulations are needed for covariance matrix analysis. Dark matter only simulations \citep[see e.g.,][and references therein]{2022LRCA....8....1A} can replicate large cosmological volumes with samples of halos and sub-halos, on top of which galaxies can be placed using a number of techniques, such as the halo occupation distribution techniques \citep[see e.g.,][]{2002ApJ...576L.105C,2002PhR...372....1C,2002ApJ...575..587B,2004ApJ...609...35K}) or the sub-halo abundance matching \citep[see e.g.,][]{2004MNRAS.353..189V,2004ApJ...609...35K,2006ApJ...647..201C,2016MNRAS.461.3421F}. In this scenario, halo properties are robustly derived from the distribution of dark matter particles \citep[see e.g.,][and references therein]{2021MNRAS.500.3309M}, and used as proxies for the generation of galaxy positions and intrinsic properties. However, dark matter only simulations can be also expensive in terms of computing time, due to the large volumes and the high mass resolution needed to accurately resolve halos and sub-halos containing the type of observed galaxies (for example, luminous red galaxies, emission-line galaxies) and again, the number of realizations demanded for robust estimates of covariance matrices. 

Approximated methods \citep[see e.g.][and references therein]{1996ApJS..103....1B, 2002MNRAS.329..629S,2013MNRAS.428.1036M,2008ApJ...678..569S,2002MNRAS.331..587M,2013MNRAS.433.2389M, 2017JCAP...07..050M,2022JCAP...11..002B,2013JCAP...06..036T,2016MNRAS.459.2118K,2018MNRAS.473.3051I,2014MNRAS.437.2594W,2015A&C....12..109H,2016MNRAS.463.2273F,2014MNRAS.439L..21K,2015MNRAS.450.1856A,2019MNRAS.483L..58B,2020MNRAS.491.2565B,2023A&A...673A.130B,2020A&A...633A..26B,2022arXiv221113590B} and learning approaches \citep[see e.g.,][]{2019arXiv190205965Z,2019PNAS..11613825H,2021ApJ...915...71V,2023MNRAS.520..668P,2024arXiv240701391D} have been shown to provide fast and relatively accurate (on scales of interest) realizations of dark matter halos with phase-space coordinates. Nevertheless, for these methods to be applicable to build galaxy mock catalogs, they need to surmount one difficulty, namely, the accomplishment of a precise assignment of halo intrinsic properties (e.g., virial mass, velocity dispersion, concentration, spin). Such task is key because the abundance and clustering probes of dark matter tracers, when explored as a function of intrinsic properties (e.g. stellar masses, X-ray luminosities) is very sensitive to the underlying scaling relations adopted to assign these properties \citep[see e.g][for an example on the construction of mock catalaogs for galaxy clusters]{2012MNRAS.425.2244B}. The assignment of halo properties constrained to reproduce the large-scale clustering signal is the scientific target of this paper, which represents the second in a number of papers dedicated to the connection between halo properties and large-scale structure \citep[][]{2023arXiv231112991B}.

Keeping in mind that $N$-body simulations can be considered to provide, to knowledge, the most realistic representation of the distribution of dark matter halos and its statistical properties in a cosmological volume, the precision of an assignment procedure can be captured in the comparison of a number of statistics, such as the halo abundance, scaling relations among the halo properties, and two (and higher) order statistics, with respect to the corresponding $N-$body estimates. In particular, the signature of large-scale effective bias \citep[see e.g.,][]{1984ApJ...284L...9K,1993ApJ...413..447F,1997MNRAS.286..795K, 10.1111/j.1365-2966.2004.07733.x, 2005MNRAS.363L..66G,2006ApJ...652...71W, 2007MNRAS.377L...5G,2007MNRAS.374.1303C, 2008MNRAS.387..921A,2008ApJ...687...12D,2010ApJ...708..469F,2017MNRAS.466.3834L,2017JCAP...03..059L,2017ApJ...848L...2M,2018MNRAS.474.5143M,2018MNRAS.476.4877M, SatoPolito2019, 2019MNRAS.484.1133C,MonteroDorta2020,MonteroDorta2021,2021MNRAS.502.3242X,2024arXiv240611182L, MonteroDorta2024} is key to connect models of galaxy/halo population with the measurements of abundance and clustering as a function of a given property. Retrieving this signal with accuracy is particularly difficult in the context of calibrated/learning oriented methods (where only phase-space coordinates are available), and the assignment of properties can only be done through the mining of the correlation between those halo properties and the underlying dark matter density field. Different approaches \citep[e.g.,][]{2015MNRAS.451.4266Z,2023A&A...673A.130B,2024MNRAS.530.2355F} have been proposed to perform this task, using local-and non-local properties on the underlying dark matter density field. Although such information helps to reproduce the clustering signal as a function of halo number counts \citep[see e.g.,][]{2019MNRAS.483L..58B}, it is not sufficient to fully replicate the signal of primary (i.e, clustering as a function of a primary halo property) and secondary large-scale bias (i.e, dependency of halo clustering on a secondary halo property at fixed values of primary property), specially for massive halos \citep[see e.g.,][]{2012ApJ...757..102W,2023A&A...673A.130B}. It is key to emphasize, in this context, that methods designed to assign galaxy properties to sub-halos based on the properties of parent halos \citep[as e.g. machine learning approaches, see e.g.,][]{2021MNRAS.503.2053R,articleForero,2022MNRAS.514.2463D, Rodrigues2023} can reproduce the large-scale bias, mainly due to the fact that such signature is already encoded in the parent halo properties used to learn from. 

In this work we present the multi-scale halo property assignment (\texttt{MSHA} hereafter) algorithm, designed to provide intrinsic properties to dark matter halos through the assessment of the scaling relation between halo properties and the statistical properties of dark matter and dark matter halos. As a novelty,  the algorithm uses the assignment of individual large-scale effective halo bias \citep[see e.g.,][]{2018MNRAS.476.3631P}, which has provided a novel  window for the analysis of primary and secondary halo bias \citep[see e.g.,][]{2021MNRAS.504.5205C,2023arXiv231112991B}, along with a multi-scale strategy to account for halo exclusion.

To assess the performance of the strategy proposed in the MSHA, we use the \texttt{UNITSim}\footnote{\url{http://www.unitsims.org/}} \citep[][]{2019MNRAS.487...48C}, which consists in \textbf{a} set of paired-fixed amplitude \citep[][]{2016MNRAS.462L...1A} cosmological $N-$body simulation, evolved from redshift $z=99$ until $z=0$ in a cosmological volume of $1({\rm Gpc} h^{-1})^{3}$ \citep[see e.g.,][for more details in this type of simulations]{2016MNRAS.462L...1A,2018ApJS..236...43G,2018ApJ...867..137V,2019MNRAS.487...48C, 2019MNRAS.490.3667Z,2020MNRAS.496.3862K, 2021MNRAS.508.4017M}. The dark matter field (constructed from $4096^{3}$ dark matter particles) is represented by a mesh of $N_{m}^{3}=256^{3}$, and halo catalogs are built using the \texttt{ROCKSTAR} algorithm \citep{2013ApJ...762..109B}, with a minimum halo-mass of $2\times 10^{11}M_{\odot} h^{-1}$ \citep[see][for more details on the halo properties in this simulation]{2023arXiv231112991B}. In forthcoming publications we shall describe ongoing developments on the assignment algorithm as well as applications to light-cones.

The structure of this paper is as follows. In \S\ref{sec:prop} we present the halo and dark matter field properties used in the \texttt{MSHA} algorithm. In \S\ref{sec:algo} we describe its main features and  present an example of the performance in \S\ref{sec:application}. We end with discussions in \S\ref{sec:dis}, and end with conclusions.

% *******************************************************************************************************************
% *******************************************************************************************************************
\section{Dark matter and dark matter halo properties}\label{sec:prop}

\subsection{Properties of the dark matter density field}
Statistical properties of the dark matter distribution $\delta_{\mathrm{dm}}(\vr)$, such as the local density or tidal field\footnote{With components $\mathcal{T}_{ij}=\partial_{i}\partial_{j}\phi$, where $\phi$ satisfies Poisson equation $\nabla^{2}\phi = \delta_{\mathrm{dm}}$} are key to reproduce halo number counts with precise two and three-point statistics \citep[see e.g.,][]{2019MNRAS.483L..58B,2020MNRAS.493..586P,2020MNRAS.491.2565B,2023A&A...673A.130B,2024arXiv240500635B}. The information of the tidal field can be used in the form of a cosmic-web classification (i.e., knots, filaments, sheets, voids; see, e.g.,  \citealt{2007MNRAS.375..489H, 2009arXiv0912.3448V, 2009MNRAS.396.1815F, 2016MNRAS.455..438A,2017ApJ...848...60Y,2018MNRAS.476.3631P}) or through the construction of combinations of its eigenvalues such as its invariants \citep[see e.g.,][]{2018MNRAS.476.3631P, 10.1093/mnras/stac671}, the prolatness, the ellipticity \citep[see e.g.,][]{2010gfe..book.....M}, the tidal anisotropy parameter, \citep[see e.g.,][]{2018MNRAS.476.3631P} or the mass of collapsing regions \citep[see e.g.,][]{2015MNRAS.451.4266Z, 2019MNRAS.483L..58B}. In this work we use the information of the local density and the cosmic-web classification as the main dark matter properties (hereafter $\{\Theta\}_{\mathrm{dm}}$), as input of the \texttt{MSHA} algorithm. The details on these properties, computed for the reference simulation, can be found at \citet[][]{2023arXiv231112991B}.

% *********************************************************************************************

% *****************************************************************
\begin{figure*}
\begin{tikzpicture}[font=\ttfamily\small,node distance=1.6cm]
\hspace{0cm}
\node(title)[title_s, xshift=8cm]{\texttt{MSHA algorithm}};
\node(Ref)[start, below of = title, yshift=0cm, xshift=-7cm]{Reference};
\node(DM)[startstop, right of=Ref, xshift=1.2cm, yshift=1cm]{Dark Matter\\ $\{\Theta_{\mathrm{dm}}\}_{\mathrm{ref}}$};
\node(H)[startstop,  right of=Ref, xshift=1.2cm, yshift=-1cm]{Halos\\ $\{\Theta_{\mathrm{H}}\}_{\mathrm{ref}} +\{\eta\}_{p}+\{\eta\}_{s} $};
\draw[arrow_ic](Ref) to [bend left =15](DM);
\draw[arrow_ic](Ref) to [bend right =15](H);
\node(HDM)[startstop,  right of=H, xshift=1.5cm, yshift=1cm]{$\{\Theta_{\mathrm{dm-H}}\}_{\mathrm{ref}}$};
\draw[arrow_ic](DM) to [bend left =15](HDM);
\draw[arrow_ic](H) to [bend right =15](HDM);
\node(P)[box_r,  below of=HDM, xshift=0cm, yshift=-1cm]{Multi-Scale};
\draw[arrow_ic](HDM) to (P);
\draw[arrow_ic](H) to (P);

\node(MS)[startstop,  below of=H, xshift=0cm, yshift=-2cm]{$\mathcal{P}_{\ell}(\{\eta\}_{p}|\{\Theta\}_{\mathrm{ref}})$\\ $\mathcal{P}(\{\eta\}_{s}|\{\eta\}_{p},\{\Theta\}_{\mathrm{ref}})$};

\node(Mock)[start, right of = Ref, xshift=14cm]{Mock};
\node(DM_M)[startstop, left of=Mock, xshift=-1.2cm, yshift=1cm]{Dark Matter\\ $\{\Theta_{\mathrm{dm}}\}_{M}$};
\node(H_M)[startstop, left of=Mock, xshift=-1.2cm, yshift=-1cm]{Halos\\ $\{\Theta_{\mathrm{H}}\}_{\mathrm{mock}}$};
\draw[arrow_ic](Mock) to [bend right =15](DM_M);
\draw[arrow_ic](Mock) to [bend left =15](H_M);
\node(HDM_M)[startstop, left of=Mock, xshift=-4.5cm, yshift=0cm]{$\{\Theta_{\mathrm{dm-H}}\}_{\mathrm{mock}}$};
\draw[arrow_ic](DM_M) to [bend right =15](HDM_M);
\draw[arrow_ic](H_M) to [bend left =15](HDM_M);

\node(TR)[box_s_k, below of = HDM_M, xshift=0cm, yshift=-3cm]{$\{\eta\}_{p}  \curvearrowleft \mathcal{P}(\{\eta\}_{p}|\{\Theta\}_{\mathrm{ref}}=\{ \Theta\}_{\mathrm{mock}}) $};
\draw[arrow_ic](HDM_M) to (TR);
%\draw[arrow_ic](P) to (TR);
\draw[arrow_ic](P) to (MS);
\draw[arrow_ic](MS) to (TR);
\node(TRs)[box_s_k, below of = TR, xshift=0cm, yshift=-0cm]{$\{\eta\}_{s}  \curvearrowleft \mathcal{P}(\{\eta\}_{s}|\{\eta\}_{p}+\{\Theta\}_{\mathrm{ref}}=\{ \Theta\}_{\mathrm{mock}}) $};
\draw[arrow_ic](TR) to (TRs);
\node(m1)[title_s, right of =TR, xshift=3cm]{Assignment of $\{\eta_{p}\}$};
\node(m2)[title_s, right of =TRs, xshift=3cm]{Assignment of $\{\eta_{s}\}$};
\end{tikzpicture}
\caption{\small{Flow-chart representing the algorithm designed to assign halo properties. The reference simulation provides a dark matter density field and its corresponding halo catalogs, with phase-space coordinates and intrinsic properties.}}    \label{fig:assig_scheme}
\end{figure*}
% *****************************************************************

\subsection{Environmental halo properties}
Following \citet[][]{2023arXiv231112991B}, we compute for each tracer a number of properties $\{\Theta\}_{\mathrm{H}}$ by collecting information of the surrounding tracers within a sphere of radius $R$, namely i) the so-called relative local mach number $\mathcal{M}_{R}$ and ii) the local overdensity $\delta_{R}$. The relative local mach number is measure of the local kinematic temperature of the tracer distribution \citep[see e.g.,][]{2001ApJ...553..513N,2013MNRAS.432..307A,2022MNRAS.512...27M}), while the local overdensity $\delta_{R}$ is defined as the excess of tracers with respect to the mean number density around each tracer. The scale $R$ used in the computation of these quantities is in principle a free parameter. We have chosen $R=3$ Mpc$/h$, a scale in which both cosmic-web environment and local overdensity help to shape the halo number clustering \citep[see e.g.,][]{2024arXiv240207995W}. The analysis by \citet[][]{2023arXiv231112991B} demonstrated that these environmental properties display the largest correlations with the large-scale effective bias. 

\subsection{Individual large-scale effective bias}
Following \citet[][]{2018MNRAS.476.3631P,2019MNRAS.489.2977R,2019MNRAS.482.1900H,  2020MNRAS.495.3233P,2021MNRAS.504.5205C,2023arXiv231112991B} we assign an effective large-scale halo bias to each tracer (with position $\vr_{i}$) as:
\be\label{eq:bias_object}
b^{(i)}_{\mathrm{eff}}=\frac{\sum_{j,k_{j}<k_{max}}N^{j}_{k}\langle {\rm e}^{-i\vk \cdot \vr_{i}} \delta_{\mathrm{dm}}^{*}(\vk) \rangle_{k_{j}}}{\sum_{j,k_{j}<k_{max}} N^{j}_{k}P_{\rm dm}(k_{j})},
\ee
where $\delta_{\mathrm{dm}}(\vk)$ is the Fourier transform of the dark matter density field, $P_{\rm dm}(k_{j})$ its power spectrum and $N^{j}_{k}$ is the number of Fourier modes in the $j$-the spherical bin in Fourier space. The sum is carried over the range of wavenumbers $k_{j}<k_{max}$ in which the ratio between the halo and the dark matter power spectra is constant \footnote{We have used a maximum wavenumber $k_{max}=0.08\,h$Mpc$^{-1}$, up to which, at the redshifts explored in this work, the ratio $P_{h}(k)/P_{dm}(k)$ is compatible with a constant value.}. 
\citet[][]{2023arXiv231112991B} showed that the effective bias as a function of multiple halo properties, as obtained from standard approaches \citep[for example, measurements of auto and cross-power halo power spectrum in bins of that halo property, see e.g.,][]{2014A&A...563A.141B,2014MNRAS.440..555P}, is consistent with the results obtained from Eq.~(\ref{eq:bias_object}) and with known calibrations in the literature, such as the bias - halo mass relation (\citealt{2010ApJ...724..878T}, see also \citealt{2018MNRAS.476.3631P}). Assigning a bias to each object allows us to include clustering signal (up to a given scale $k_{\max}$) to the machinery for assigning halo properties. Notice that we can also compute secondary bias \citep[see e.g.,][]{2009PhRvD..80f3528S}, which can help to improve the signal of higher order statistics in terms of halo properties. For the purpose of this work, we only use the effective bias and leave quadratic bias for a forthcoming research.

Compared with previous attempts to assign halo properties, the inclusion of the large-scale effective bias is key to retrieve properties that replicate the large-scale pattern as drawn by such tracers. As we will show in the following sections, not including this bias yields halo properties which, even if depicting the correct abundance, fail, particularly at the high mass end of the halo mass function, at reproducing primary and secondary bias, even if non-local properties of the dark matter field are included.

\subsection{Target halo properties}\label{sec:target}
Let us briefly describe the halo properties we will aim at reconstructing. In general, the properties of dark matter halos can be separated in a primary $\eta_{p}$ and secondary $\eta_{s}$ set. The definition of these sets can be established by a number of ways, such as correlation with the underlying dark matter density field \citep[see e.g.,][]{2023A&A...673A.130B} or a principal component analysis of the halo properties \citep[][]{2011MNRAS.416.2388S,2023MNRAS.523.5789Z,2023arXiv231112991B}. We can call ``primary" properties as those directly probing the depth of the potential wells of dark matter halos, such as virial mass, velocity dispersion or maximum circular velocity $V_{\mmax}$ \citep[see e.g.,][]{2019ApJ...887...17Z}. In this work we use the latter  as main property, followed by the virial mass $M_{\vvir}$. As secondary, we use halo spin \cite[][]{2001ApJ...555..240B} and the halo concentration $C_{\vvir}\equiv R_{\vvir}/R_{s}$, where $R_{\vvir}$ is the virial radius and $R_{s}$ the scale radius, obtained based on fits to a NFW profile \citep[]{1996ApJ...462..563N} within the \texttt{ROCKSTAR} algorithm \citep[see e.g.,][]{2011MNRAS.415.2293K,2020MNRAS.493.4763M}.

% *****************************************************************
\begin{figure*}
\includegraphics[trim =.26cm 0.25cm .2cm 0cm ,clip=true, width=0.33\textwidth]{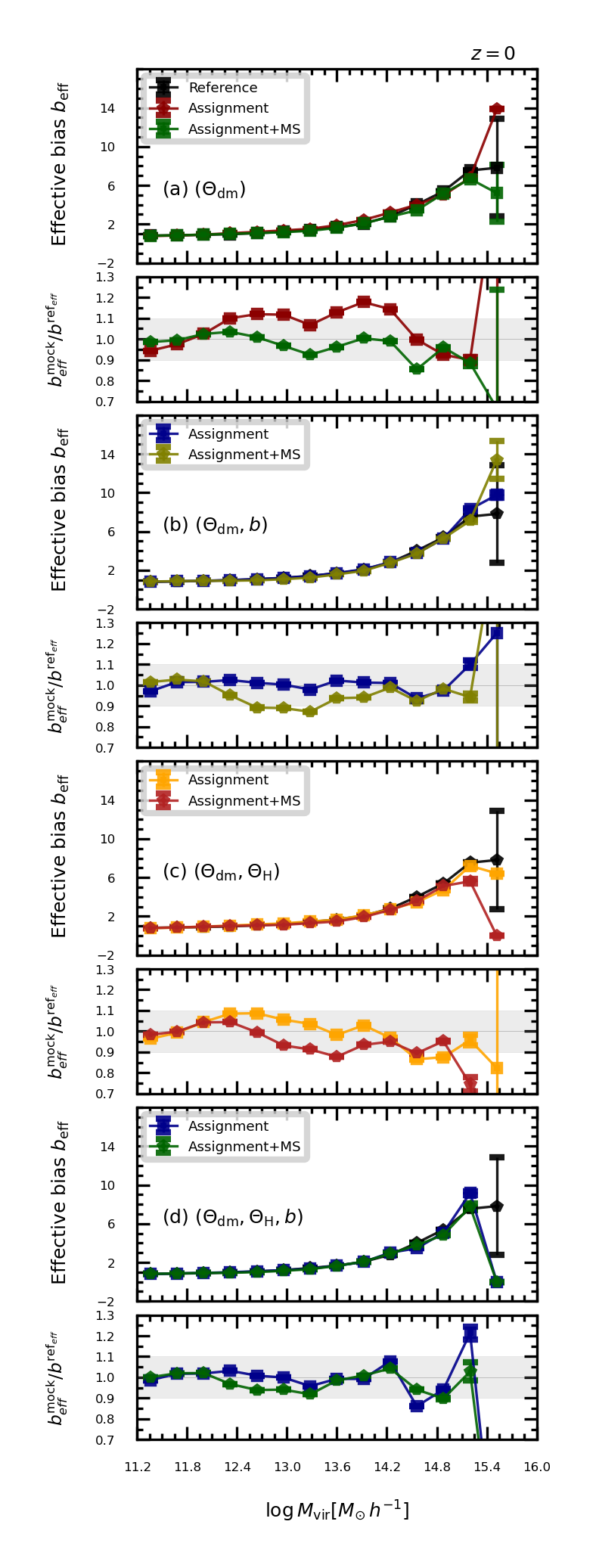}
\includegraphics[trim =.26cm 0.25cm .2cm 0cm ,clip=true, width=0.33\textwidth]{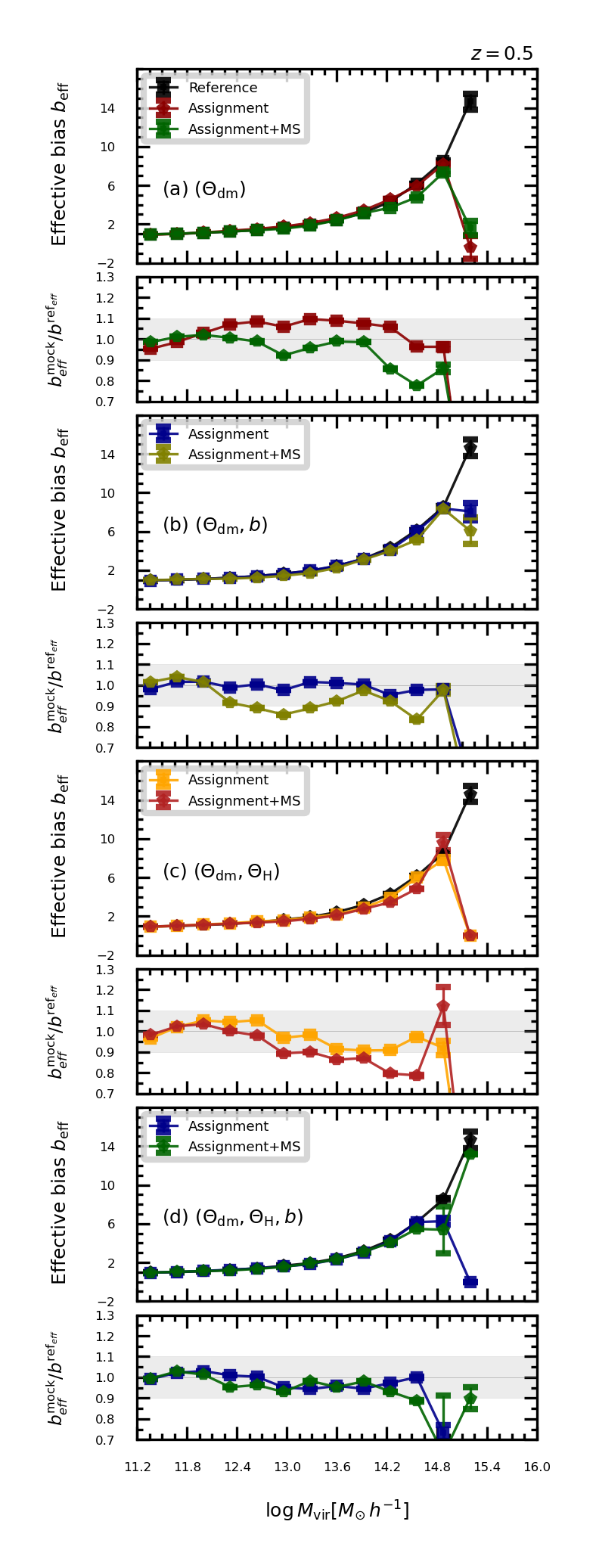}
\includegraphics[trim =.26cm 0.25cm .2cm 0cm ,clip=true, width=0.33\textwidth]{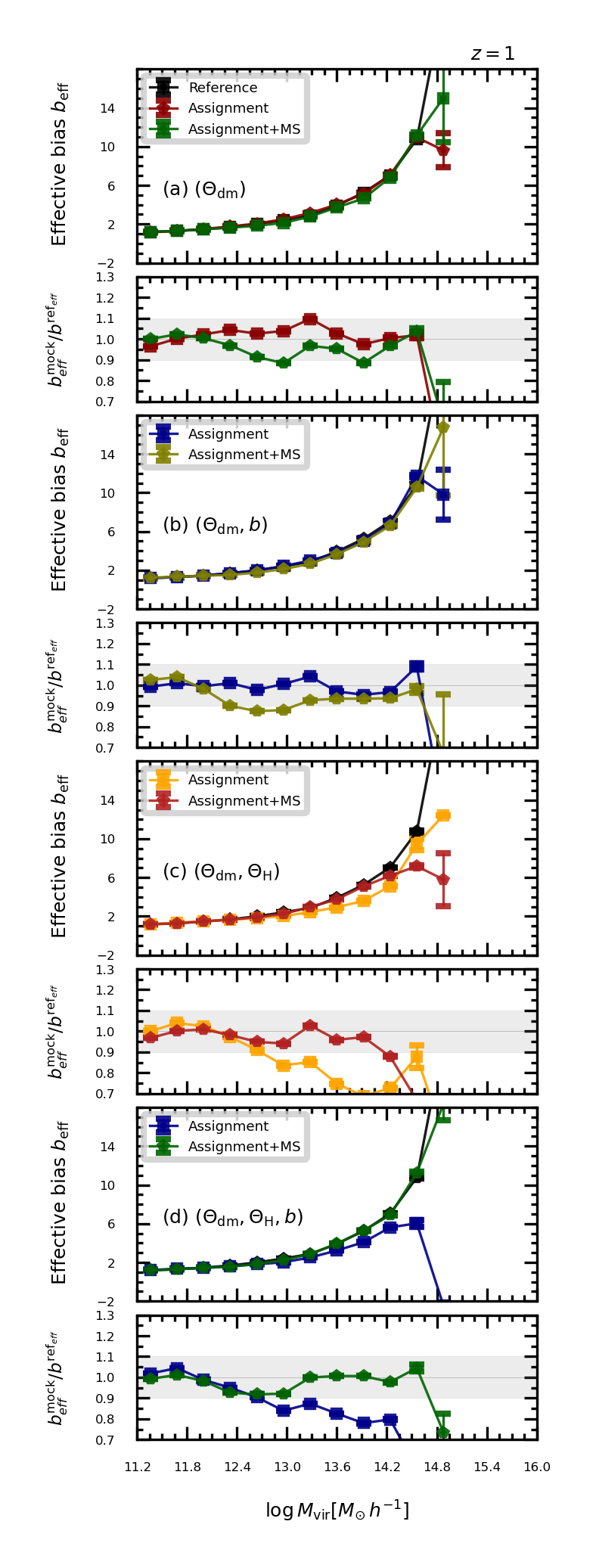}

\caption{\small{Large-scale effective bias as a function of virial mass, assigned as described in \S\ref{sec:algo}. The main panels (a,b,c,d) show the effective bias measured from the reference and the results from the property assignment at $z=0$ (left) and $z=0.5$ (center) and $z=1$ (right column), with (labeled Assignment$+$MS) and without (labeled Assignment) the multi-scale approach. Panel (a) shows the results of using only dark matter properties $\Theta_{\mathrm{dm}}$ in the assignment of maximum circular velocity and virial mass. Panel (b) shows the results of adding the information from the individual bias. Panel (c) shows the results after including halo environmental properties $\Theta_{\mathrm{H}}$, and panel (d) shows the result of including both bias and environmental properties. The bottom panels show, in each case, the ratio to the primary bias obtained from the mock catalog using its original properties (the latter shown as black points in all main panels). The gray region in these bottom panels depicts $10\%$ deviation with respect to unity. The error bars denote the error on the mean bias-property relation.}}
\label{fig:vmax1}
\end{figure*}
% *****************************************************************

% **************************************************************************************
% **************************************************************************************
\section{The algorithm}\label{sec:algo}

The \texttt{MSHA} algorithm \footnote{The \texttt{MSHA} algorithm is part of the \texttt{C++} library \texttt{CosmiCodes} found at \url{https://github.com/balaguera/CosmicCodes}.} aims at assigning halo intrinsic properties (as those discussed in \S\ref{sec:target}) constrained to provide the correct large-scale clustering signal as a function of those properties. The input of the algorithm is made up by two components, namely:
\begin{itemize}
    \item A reference simulation, which is ideally one (or a number of) realization(s) of a tracer distribution in an $N$-body simulation, from which we can have access to dark matter density field and a halo catalog (with intrinsic properties).
    \item  A target or mock realization of dark matter halos (with phase space coordinates) together with its corresponding dark matter density field. 
\end{itemize}
Two main features characterize this algorithm. On one hand, we explicitly implement the assignment of individual halo bias of Eq.~(\ref{eq:bias_object}) to mine its correlations with halo properties \citep[][]{2023arXiv231112991B}. This is a key ingredient, as it directs the distribution of halo properties towards the correct large-scale signal when the two-point statistics of tracers is assessed as a function of their intrinsic properties. On the other hand, \texttt{MSHA} accounts for halo-exclusion \citep[see e.g.,][]{2002ApJ...565...24P,2014A&A...563A.141B,2019MNRAS.489.4170G} by implementing a so-called multi-scale algorithm (MS hereafter) described by \citet[][]{2023A&A...673A.130B}. 

The main idea behind the MS approach is that tracers with largest values of primary properties $\eta_{p}$ (e.g. virial mass, maximum circular velocity or velocity dispersion) are those delineating the statistical properties on large-scales. Accordingly, high values of such properties should be prevented from being assigned to close pairs. To achieve this, \texttt{MSHA} divides the range of available values of the reference primary property $\eta_{p}$ in a number of intervals or levels. Each level is in turn characterised by a spatial mesh covering the simulation volume, whose resolution is given by the number of reference tracers with values $\eta_{p}$ in that level, such that on average, each spatial cell within each level is populated by one tracer \footnote{The algorithm uses as input parameter the number of levels and the size of the mesh of each level, from which the range of the intervals in the property $\eta_{p}$ are computed. That is, the number of tracers in each interval equals the size of the mesh given as input parameter.}. The exclusion effect, being linked to the size of dark matter halos, can be more pronounced towards low redshift, with the increment in the abundance of high-mass halos \citep[see e.g.][]{2014A&A...563A.141B}.

% *****************************************************************
\begin{figure*}
\includegraphics[trim =.26cm 0.2cm .2cm 0cm ,clip=true, width=0.33\textwidth]{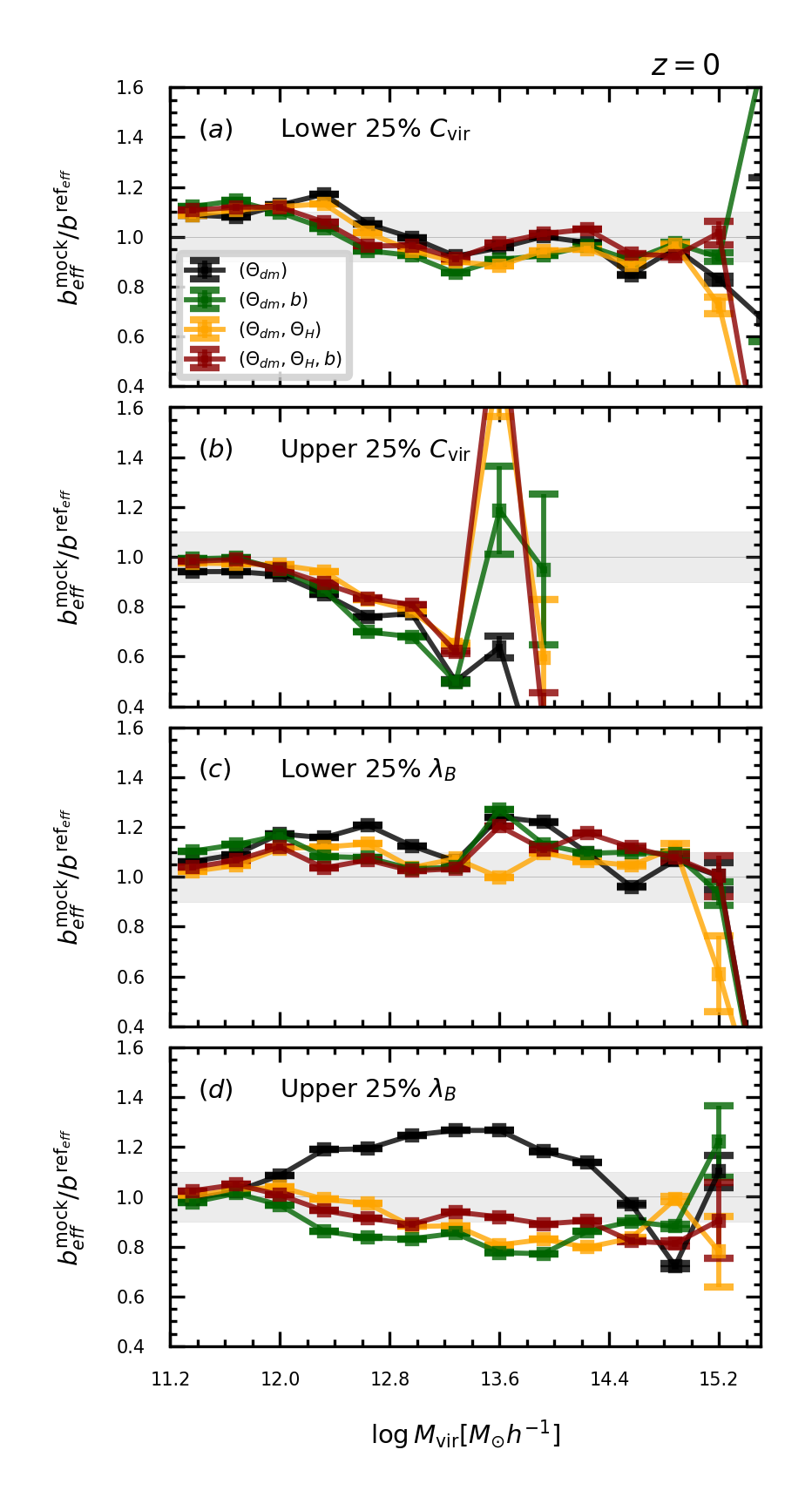}
\includegraphics[trim =.26cm 0.2cm .2cm 0cm ,clip=true, width=0.33\textwidth]{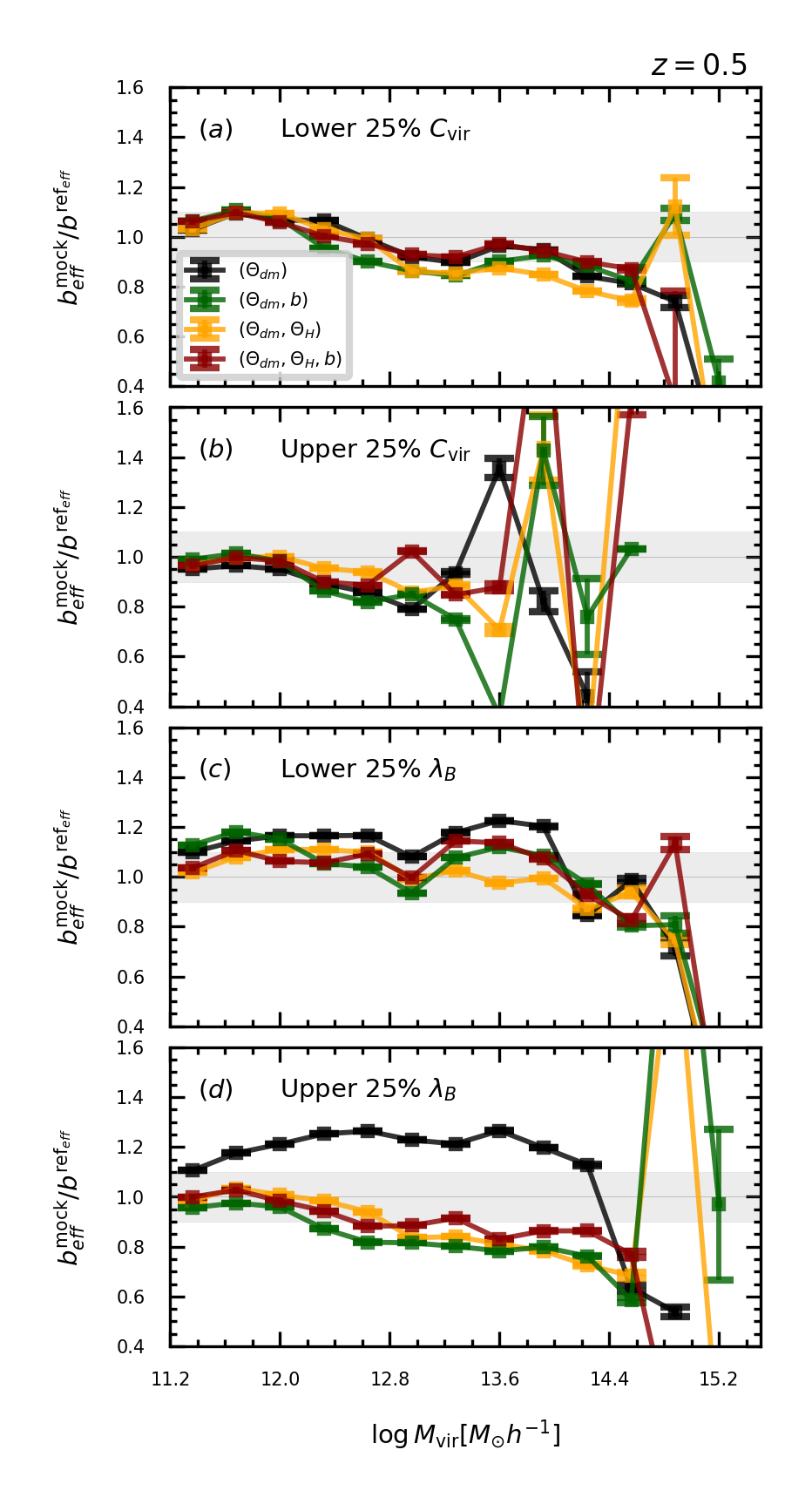}
\includegraphics[trim =.26cm 0.2cm .2cm 0cm ,clip=true, width=0.33\textwidth]{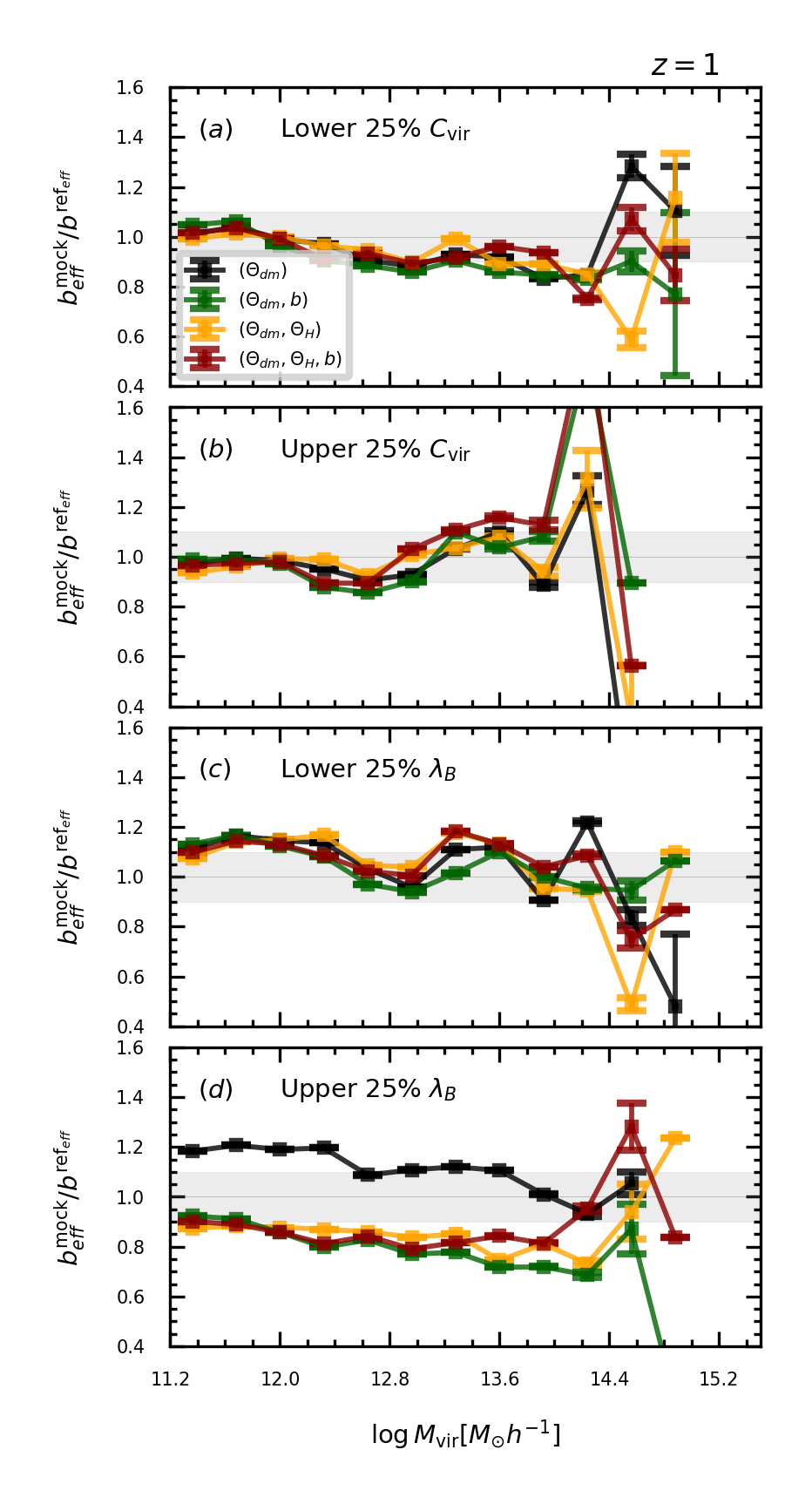}
\caption{\small{Secondary effective bias as a function of halo virial mass. This plot compares (with ratios) the effective bias measured reference and the mock after assignment, at two different cosmological redshifts (shown in different columns) and using different combinations of halo/dark matter properties, as shown in the legends. Panels (a) and (b) show the secondary bias using halo concentration, for the lower and upper quartiles computed in each mass bin, respectively. Similarly, panels (c) and (d) shows secondary bias as a function of halo spin. In all plots the shaded region denotes a $10\%$ deviation from unity. In all cases, MS is employed. The error bars are computed from adding in quadrature the corresponding uncertainties in each bias estimate.}}
\label{fig:vmax3a}
\end{figure*}
% *****************************************************************

The assignment of halo properties is performed in a hierarchical order, in which a main property is assigned first (in our case, $V_{\mathrm{max}}$) and subsequent assignment campaigns follow using the information of the properties already assigned \citep[see e.g.,][for applications to hydro-simulations]{2021ApJ...921...66S}. The first assignment is done explicitly using the MS technique, reading the values of the main property from the reference catalog, and assigning these to mock tracers in a top-bottom order: that is, starting from the highest level (containing the highest values of $\eta_{p}$), it sorts the values of the properties $\eta_{p}$ and assigns them from highest values in each level, simultaneously preventing from assigning two consecutive values of $\eta_{p}$ to tracers living in the same spatial cell. 
Values below the minimum of $\eta_{p}$ defined by the last level are assigned from the (unsorted) remaining set of reference properties, which is indeed the approach followed for all mock tracers when MS is not applied. In both cases, the process takes into account the dependencies with local and non-local properties $\{\Theta\}$ presented in \S\ref{sec:prop}. 

The \texttt{MSHA} algorithm contains a number of options to prevent failed attempts to assign the main property. The main source of failure is linked to the so-called cosmic-variance, which is embodied in this scenario as the difference in the actual number of tracers in the reference and tracers in the mock in a given bin of $\{\Theta\}$. To mitigate this, the MS approach uses tolerance percentages $f_{\ell}\leq 1$ to reduce the number of tracers to be assigned in a given level, assigning the remaining fraction as in a non-MS scenario. In that case, if the number of requested properties is larger than the number of available properties in the reference (in a given bin of $\{\Theta\}$), the property is assigned following the conditional probability distribution $\mathcal{P}_{\ell}(\eta_{p}|\{\Theta\})$ assessed for every level $\ell$. Finally, if the set of properties $\{\Theta\}_{\mathrm{mock}}$ find no representation in the set $\{\Theta\}_{\mathrm{ref}}$, the value of $\eta_{p}$ assigned to the tracers is drawn from the global distribution.  
%$\mathcal{P}(\eta_{p})=\sum_{\ell}\int \mathrm{d}\Theta \mathcal{P}_{\ell}(\eta_{p}|\{\Theta\})$. 

The flow-chart of Fig.~\ref{fig:assig_scheme} depicts the main steps followed in the assignment of halo properties. This procedure can be generalized to any number of reference simulations (strictly speaking, an ensemble of independent realizations built from the same initial power spectrum) from which the conditional probability distributions $\mathcal{P}(\eta_{p}|\{\Theta\})$ can be obtained \citep[as shown by][]{2023A&A...673A.130B}.

% *******************************************************************************************************************
\section{Application}\label{sec:application}
We use a set of paired and fixed-amplitude $N$-body simulations (at redshift $z=0,0.5$ and $z=1$) to show the performance of the method. This represents the set reference-mock. With this type of simulations, even if the initial conditions display the same amplitude of power spectrum, differences in the occupancy of the bins designed to build the scaling relations can arise through gravitational evolution-coupling of initially randomly distributed phases \citep[see e.g.,][]{2016MNRAS.462L...1A,2018ApJ...867..137V}.

Following the procedure described in \S\ref{sec:algo}, we assign maximum circular velocities to the halos in the mock simulation on top of which we assign virial mass\footnote{For the assignment of $V_{\mathrm{max}}$ using MS, we have used three levels with sizes $16^{3},64^{3},128^{3}$, with $100$ bins in local density, $4$ cosmic-web types, $10$ bins in local overdensity, $10$ bins in Mach number and $15$ bins in large-scale bias. The tolerance factors are chosen to be $0.9$ for all levels. At $z=0$, the fraction of $V_{\mathrm{max}}$ assigned in each level is $\sim 0.01, 1.2$ and $\sim 10\%$ of the total number of available tracers, for levels $\ell=1$ (most massive halos), $\ell=2$ and $\ell=3$ respectively. Below the last level, no MS is applied; in this case, $\sim 70\%$ of tracers are assigned a property from the reference, while the remaining fraction are assigned a property using the global distribution. For the assignment of virial mass and secondary properties, $100$ bins in $V_{\mathrm{max}}$ and $M_{\vvir}$ are used, respectively. This set-up leads to $\sim 3,7,8$ and $12\%$ global differences in abundance with respect to the reference, when measured as a function of $V_{\mathrm{max}}$, $M_{\vvir}$, $C_{\vvir}$ and $\lambda_{B}$ respectively.} through the assessment of the probability distribution $\mathcal{P}(M^{\mathrm{ref}}_{\vvir}|V^{\mathrm{ref}}_{\rm max}, \{\Theta\}_{\mathrm{ref}})$, sampling as:
\be\label{eq:mass_assi}
M_{\vvir}  \curvearrowleft \mathcal{P}\lp M_{\vvir}^{\mathrm{ref}}  \mid \, V^{i}_{\mmax} \in \Delta V_{\mmax}^{\mathrm{ref}}, \{\Theta\}_{\mathrm{ref}}=\{\Theta\}_{\mathrm{mock}} \rp  .
\ee                
In Fig.\ref{fig:vmax1} we present the impact of the property assignment in the large-scale effective bias as a function of virial mass\footnote{Even if the first assignment corresponds to $V_{\mathrm{max}}$, we show the results in terms of $M_{\mathrm{vir}}$ as it is more familiar within the large-scale structure community}. In particular, we show three different scenarios, namely i) assignment using only dark matter properties, ii) adding the effective bias, iii) using dark matter and halo environmental properties and iv) adding the effective bias the previous set, for three cosmological redshifts ($z=0,0.5,1$). In each of these cases, we present the results with and without the MS approach applied in the assignment of $V_{\mmax}$. 

In general, Fig.~\ref{fig:vmax1} shows that the implementation of the information of halo environmental properties and the large-scale bias yields a clustering signal in agreement, on large scales, with the clustering signal measured from the  mock catalog using the original properties, especially on the high mass halo population (where the impact of the MS approach is noticeable, given that exclusion effects are more dominant as long as $z\to 0$ and at large scales). It also is observed that as long as we approach $z=0$, the presence for properties beyond those of the dark matter is more evident (we have verified this trend performing the same analysis at $z=0.5$ and $z=3$).  

We now use halo concentration $C_{\vvir}$ and halo dimensionless spin $\lambda_{B}$ as example of the assignment of secondary properties. To establish the hierarchical approach to the assignment of properties, we refer to \citet[][]{2023arXiv231112991B}, where it was shown that the correlation between halo concentration and mass is larger than that between mass and halo spin. Accordingly, we start by assigning halo concentration by measuring from the reference the scaling relation $\mathcal{P}(C_{\vvir} |M^{\mathrm{ref}}_{\vvir},\{\Theta\}_{\mathrm{ref}})$ and sampling from it as 
\be\label{eq:cvir_assi}
C_{\vvir}  \curvearrowleft \mathcal{P}\lp C_{\vvir}^{\mathrm{ref}} \mid  M^{i}_{\vvir} \in \Delta M_{\vvir}^{\mathrm{ref}} , \{\Theta\}_{\mathrm{ref}}=\{\Theta\}_{\mathrm{mock}} \rp,
\ee                
where the masses $M^{i}_{\vvir}$ correspond to the values assigned with Eq.~(\ref{eq:mass_assi}). Similarly, for the assignment of halo spin, we measure the scaling relation $\mathcal{P}(\lambda_{B} |M^{\mathrm{ref}}_{\vvir}, C^{\mathrm{ref}}_{\vvir},\{\Theta\}_{\mathrm{H,ref}}, \{\Theta_{H} ) $ and sample values as
\be\label{eq:spin_assi}
\lambda_{B}  \curvearrowleft \mathcal{P}\lp \lambda_{B}^{\mathrm{ref}} \mid  M^{i}_{\vvir} \in \Delta M_{\vvir}^{\mathrm{ref}} , C^{i}_{\vvir} \in \Delta C_{\vvir}^{\mathrm{ref}} ,\Theta_{\mathrm{ref},H}=\Theta_{\mathrm{mock},H}  \rp,
\ee                
where we have not included the dark matter properties, and solely relied on the halo intrinsic and environmental properties.

To assess the precision of the the signal of secondary bias, we divide the sample in quartiles of the secondary property and measure the mean effective bias in bins of a primary property. In Fig.~\ref{fig:vmax3a} we display the ratios of effective bias in these two quartiles of the secondary properties with respect to the same signal measured from the reference simulations, and for three different snapshots of the \texttt{UNITSim}. In general, all tests display $\sim 10\%$ discrepancy with respect to the reference. The largest difference is obtained when only dark matter properties are used, in agreement with previous results in the literature. The inclusion of halo environmental properties $\Theta_{H}$ and/or effective bias contribute to assign halo properties with a secondary bias signature within $\sim 10\%$ difference with respect to the reference. The results of Fig.~\ref{fig:vmax3a} have been obtained using MS, and we have verified that such choice produce better results than the case without MS at the redshifts \textbf{shown in that figure}.

\section{Discussion}\label{sec:dis}
Our results show that the use of individual effective bias (Eq.~\ref{eq:bias_object}) can be relevant in the process of reconstructing and/or assigning halo properties to generate reliable halo mock catalogs that follow the expected large-scale clustering signal. While the absence of the information on effective bias or environmental halo properties in the assignment can still generate robust scaling relations between halo properties (specially towards high redshifts, as seen in panels (a) of Fig.\ref{fig:vmax1}), their use can improve the resulting clustering signal. 
The signal of mean large-scale primary bias (i.e, as shown in Fig. \ref{fig:vmax1}) is one of the key probes in the clustering analysis of current galaxy redshift surveys \citep[see e.g.,][]{2020A&A...643A..70T}, as it can provide insights on the galaxy/cluster scaling relations \citep[see e.g.,][]{2002PhR...372....1C,2014A&A...563A.141B,2018MNRAS.473.2486S} and physics of the early Universe \citep[i.e, primordial non-Gaussianities, see e.g.,][]{2012MNRAS.422.2854G,2023arXiv231015731E}. As such, it is important to minimize deviations in the mean primary bias (with respect to the results of $N$-body simulations) when constructing mock catalogs,  as it can propagate from the covariance matrix to the likelihood and lead to biased results \citep[][]{2021A&A...649A..47A,2023A&A...671A.100E}
% *****************************************************************
\begin{figure}
\includegraphics[trim =.2cm 0.2cm 0cm 0cm ,clip=true, width=0.5\textwidth]{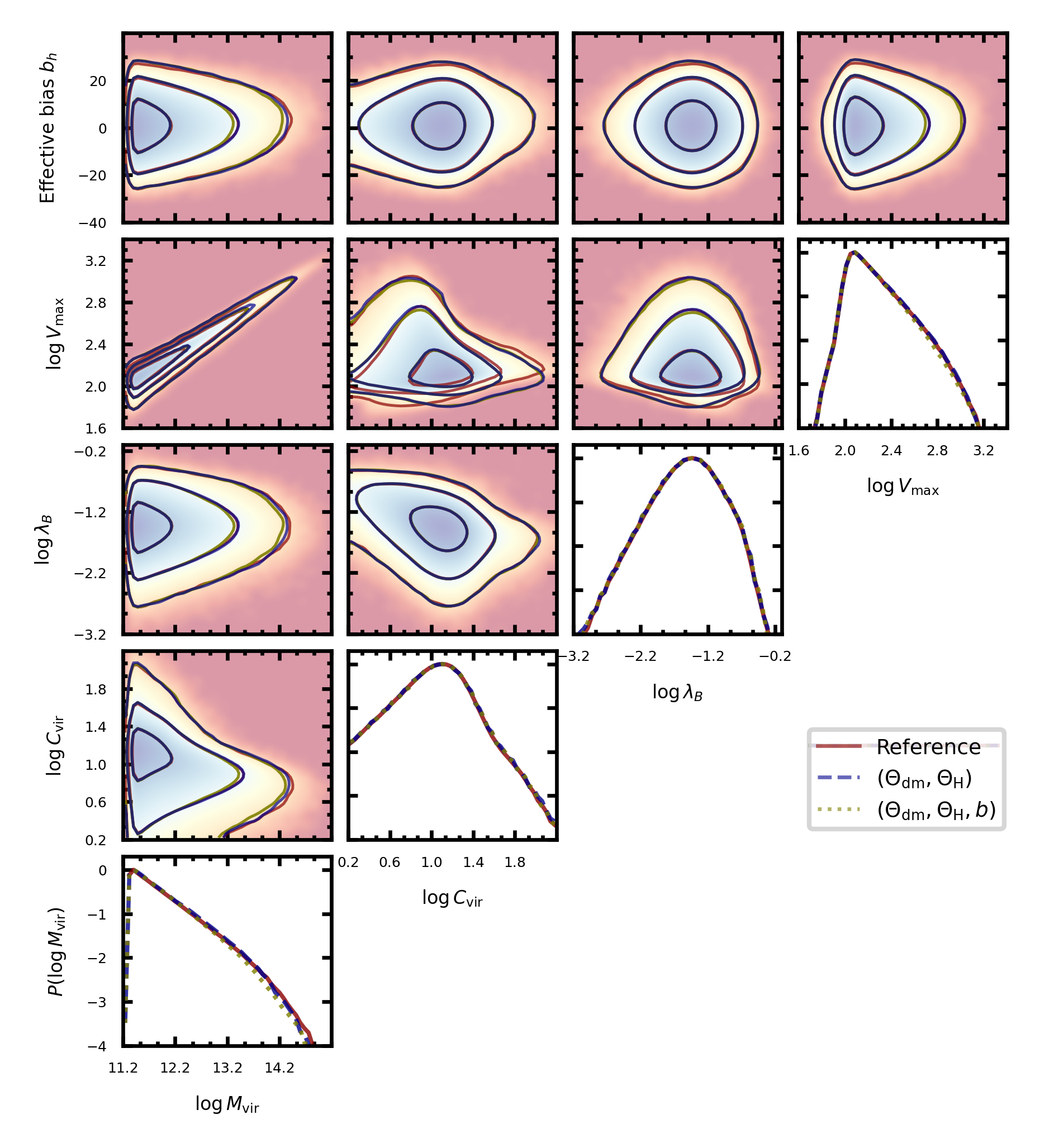}
\caption{\small{2D scaling relations between halo properties and large-scale halo effective bias in the \texttt{UNITSim} at $z=0$. The different lines in the contours (denoting surfaces of equal number of tracers) and the 1D distributions show the result from different set-ups in the assignment algorithm, as shown in the legend.}}
\label{fig:sr}
\end{figure}
% *****************************************************************

The inclusion of environmental halo properties and effective bias is also key (to a evident degree though) if secondary bias (key for assessing the impact of e.g., HOD assumptions on the galaxy population, see e.g., \citet[][]{2016MNRAS.460.2552H,2017ApJ...834...37L}) is expected to be present in the mock catalog. It can be shown that effective bias as a function of, e.g, virial mass, in quartiles $\Delta \eta_{s}=\Delta \eta_{s}(M_{\vvir})$ of a secondary property $\eta_{s}$ can be mathematically decomposed as 
\be
\langle b|M_{\vvir}\rangle_{\Delta\eta_{s}}=\int \mathrm{d}\tilde{M}_{\vvir}\langle b|\tilde{M}_{\vvir}\rangle F_{\Delta\eta_{s}}(M_{\vvir},\tilde{M}_{\vvir}),
\ee
where $\langle b|\tilde{M}_{\vvir}\rangle$ is the primary bias,
\be
F_{\eta_{s}}(M_{\vvir},\tilde{M}_{\vvir})\equiv \int_{\Delta \eta_{s}} \mathrm{d}\eta_{s}\mathcal{P}(\eta_{s}|M_{\vvir})\mathcal{P}(\tilde{M}_{\vvir}|\eta_{s}),
\ee
and $\mathcal{P}(\eta_{s}|M_{\vvir})$ denotes the scaling relation between $\eta_{p}$ and $M_{\vvir}$. Thus, the signal of secondary bias (e.g., as shown in Fig.~\ref{fig:vmax3a}) is sensitive to the full scaling relation (i.e, all the moments of the conditional probability distribution $\mathcal{P}(\eta_{s}|M_{\vvir})$ In Fig.~\ref{fig:sr} we show the 2D and 1D distribution of halo properties and effective bias from the reference simulation as well as from the different experiments performed in this paper. While there are sizable differences (for example, in the $V_{\mathrm{max}}-C_{\vvir}$ relation), in general, the reconstruction of the halo properties provides not only a robust set of halo-scaling relations, but also reconstructs the large-scale halo connection through primary ($\leq 5\%$)  and secondary ($\leq 15\%$ differences) bias, as can be read from Figs.~\ref{fig:vmax3a} and \ref{fig:sr}.

The performance of the \texttt{MSHA} algorithm can depend on how detailed the different scaling relations (assessed from the reference and to be sampled onto the mock catalog) are determined. However, the level of refinement is not arbitrary as we can eventually enter an overfitting regime. This can eventually precipitate a failed attempt to assign properties when a new set of density field (or another realization of the reference fields) and its halo distribution is used as a target mock and, as such, a limit in the number of bins (or tolerance thresholds) is imposed.

\begin{figure}
\includegraphics[trim =.2cm 0.2cm 0cm 0cm ,clip=true, width=0.5\textwidth]{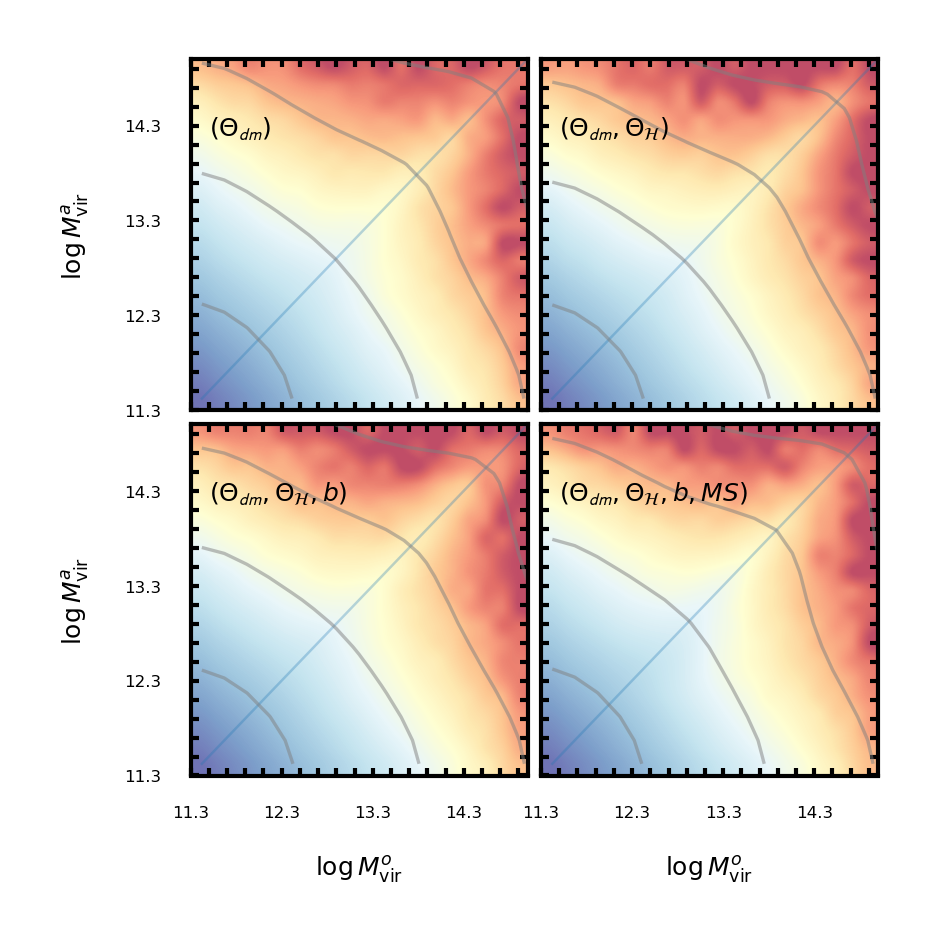}
\caption{\small{Scaling relation between the logarithm of the assigned virial mass $M_{\mathrm{vir}}^{a}$ and the logarithm of the true virial mass $M_{\mathrm{vir}}^{o}$}, for four different set-ups in the assignment algorithm (at $z=0$). The color coding and contours indicate constant number of tracers from high (blue) to low (red).}
\label{fig:rec}
\end{figure}

\begin{figure}
\includegraphics[trim =.2cm 0.25cm 0cm 0cm ,clip=true, width=0.5\textwidth]{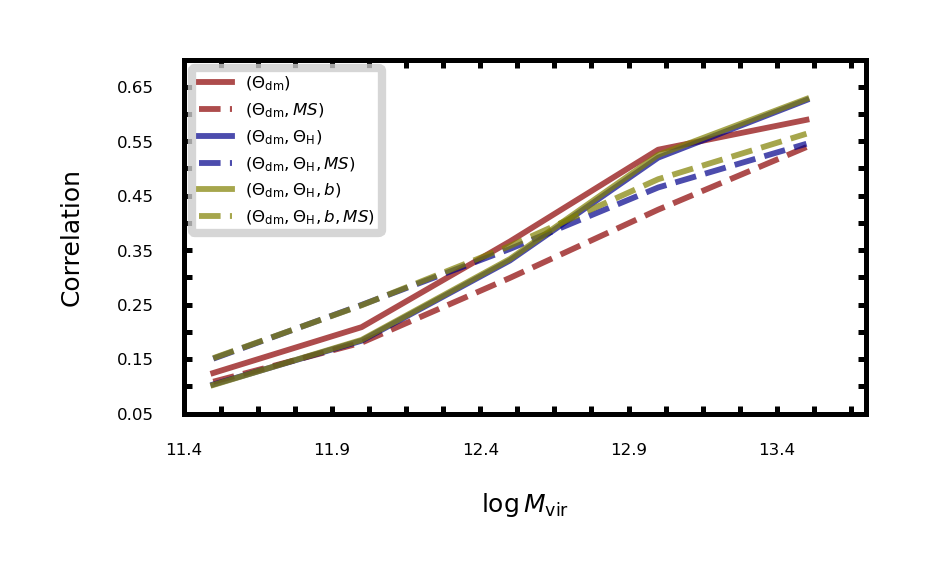}
\caption{\small{Correlation coefficient between the assigned mass and the original mass at $z=0$ for different set-ups used in the assignment algorithm, for cuts in the true and assigned masses $M^{a}_{\mathrm{vir}},M^{0}_{\mathrm{vir}}>M_{\mathrm{vir}}$}}
\label{fig:re_m}
\end{figure}

In Fig.~\ref{fig:rec} we present an example of the scaling relation between the assigned halo mass and the true value of that property as read from the reference simulation used as mock catalog (at $z=0$). In general, there is a large scatter between the true and assigned property, which varies little among the different set-ups shown. In Fig.~\ref{fig:re_m} we display the correlation coefficient between the properties shown in Fig.~\ref{fig:rec} for halos with true and assigned masses above the values in the x-axis, indicating that in general the reconstruction improves towards large halo masses. The large scatter seen at the low halo mass region likely arises from the degeneracies between the different dark matter properties (e.g. cosmic-web types) and the mass of the halos hosted therein, which cannot be broken by the inclusion of large scale bias \citep[as it also shows large scatter with respect to halo mass, see e.g.,][]{2023arXiv231112991B}. While the impact of such degeneracy is small when measuring the signal of large-scale effective bias, it can be crucial when populating intermediate and low mass halos with galaxies. In forthcoming work we will add learning algorithms to reduce this scatter and provide a more accurate reconstruction of halo properties.

% *****************************************************************
% *****************************************************************
% *****************************************************************

% *******************************************************************************************************************
% *******************************************************************************************************************
\section{Conclusions}
In this paper we have described a strategy to assign intrinsic halo properties in halo mock catalogs generated with the so-called calibrated methods \citep[see e.g.,][]{2019MNRAS.483L..58B}. These methods can provide sets of independent halo realizations with phase-space coordinates and their underlying dark matter density field, lacking though of intrinsic halo properties, which are key in order to generate galaxy mock catalogs using, for example, an HOD approach \citep[see e.g.,][]{2013MNRAS.433.2389M,2020MNRAS.497..581A,2024arXiv240513495E}.

While the assignment of halo properties can be accurately performed based on the abundance or conditional probability distributions, the resulting two-point statistics on large-scales, expressed as a function of that particular property, is not fully captured \citep[see for example][]{2015MNRAS.451.4266Z, 2023A&A...673A.130B}. A higher precision in the clustering signal (with respect to $N-$body estimates) can be achieved when conditional mass (or any other property) function are implemented, using local (i.e, density) and non-local (i.e, tidal field) information of the underlying dark matter density field. Nevertheless, such ingredients cannot induce the correct signal of large-scale effective bias, specially for massive tracers.

With the possibility of individually computing large-scale bias for dark matter tracers \citep[see e.g.,][]{2018MNRAS.476.3631P, 2021MNRAS.504.5205C, 2023arXiv231112991B}, it is feasible to assign intrinsic halo properties ensuring that the two-point statistics, in particular, the signal of primary and secondary halo bias  measured on large-scales is closer to what is measured from a, e.g., $N-$body simulation. The algorithm described in this paper makes use of this fact, complementing it with with a so-called multi-scaling approach, designed to account for the phenomena of halo exclusion. It can implement other dark matter properties, such as tidal anisotropy, peak statistics \citep[see e.g.,][]{Peacock1985} or the shear velocity field \citep[see e.g.,][]{2018MNRAS.473.1195L}. Similarly, other non-local halo properties such as the neighbor statistics can be included.

We can summarize the main results of this paper as follows:
\begin{itemize}
    \item Using the information of the halo individual large-scale bias, environmental halo and dark matter properties plus the multi-scale approach, the \texttt{MSHA} algorithm can assign intrinsic properties to dark matter halos with a resulting large-scale primary bias (i.e., expressed as a function of primary properties such as virial mass or maximum circular velocity) within $\sim 5\%$ precision with respect to the signal obtained using the properties in the the reference simulation (see Fig.~\ref{fig:vmax1}). 
    \item The reconstruction of the secondary bias is also favored by the inclusion of the large-scale individual bias, although the differences among the different set-ups shown in Fig.~\ref{fig:vmax3a} are rather small.
    \item As a byproduct of the reconstruction of halo bias, halo properties, such as virial mas, are also recovered. This yields correlations between assigned and original halo properties of $\sim 0.5$ for high masses, as shown in Fig.~\ref{fig:rec}. We note that the algorithm in its current state is not specifically designed to recover halo properties with high precision, but to replicate the property-bias relation. 
   
    \item Our results motivate further developments of this algorithm, using, e.g., learning techniques, that helps to improve the precision in the reconstruction of intrinsic properties, especially for low mass halos, which is key to the assignment of galaxies to dark matter halos in mock catalogs generated with approximated methods. 
\end{itemize}

\begin{acknowledgements}
 We would like to thank the referee for her/his review, which helped in the presentation of our results. We would like to thank J.Garcia-Farieta for comments on the manuscript. We also acknowledge Chia-Hsun Chuang, Gustavo Yépez and F.-S. Kitaura for granting access to the \texttt{UNITSim}. ABA acknowledges the Spanish Ministry of Economy and Competitiveness (MINECO) under the Severo Ochoa program SEV-2015-0548 grants. The \texttt{UNITSim} has been run at the MareNostrum Supercomputer hosted by the Barcelona Supercomputing Center, Spain, with computing time granted by PRICE under grant number 2016163937. ADMD thanks the IAC facilities and Fondecyt for financial support through the Fondecyt Regular 2021 grant 1210612.  No AI application has been used to generate neither text, figures data, nor code for this article.
\end{acknowledgements}

\bibliographystyle{aa}
\bibliography{refs}  

\end{document}